\documentclass{sig-alternate-05-2015}
\usepackage{booktabs} 
\usepackage{subcaption}
\begin{document}
\title{Evaluating Memento Service Optimizations}
\numberofauthors{3}
\author{
%
\alignauthor
Martin Klein \\
        \affaddr{Research Library}\\
        \affaddr{Los Alamos National Laboratory}\\
        \affaddr{Los Alamos, NM, USA}\\
        \affaddr{\url{http://orcid.org/0000-0003-0130-2097}}\\
        \email{mklein@lanl.gov}
\and
\alignauthor
Lyudmila Balakireva\\
        \affaddr{Research Library}\\
        \affaddr{Los Alamos National Laboratory}\\
        \affaddr{Los Alamos, NM, USA}\\
        \affaddr{\url{http://orcid.org/0000-0002-3919-3634}}\\
        \email{ludab@lanl.gov}
\alignauthor
Harihar Shankar\\
        \affaddr{Research Library}\\
        \affaddr{Los Alamos National Laboratory}\\
        \affaddr{Los Alamos, NM, USA}\\
        \affaddr{\url{http://orcid.org/0000-0003-4949-0728}}\\
        \email{harihar@lanl.gov}
}
%
%
%
%
\maketitle
\begin{abstract}
Services and applications based on the Memento Aggregator can suffer from slow response times due to the federated search across
web archives performed by the Memento infrastructure. In an effort to decrease the response times, we established a cache system 
and experimented with machine learning models to predict archival holdings. We reported on the experimental results in previous 
work and can now, after these optimizations have been in production for two years, evaluate their efficiency, based on long-term
log data. During our investigation we find that the cache is very effective with a $70-80\%$ cache hit rate for human-driven 
services. The machine learning prediction operates at an acceptable average recall level of $0.727$ but our results also show that 
a more frequent retraining of the models is needed to further improve prediction accuracy.
\end{abstract}
%
%
%
%
%
%
\maketitle
\section{Introduction}
Since Memento was standardized in $2013$ \cite{memento}, a variety of services have emerged based on the protocol. Examples are:
\begin{itemize}
\item the Memento TimeTravel web service\footnote{\url{http://timetravel.mementoweb.org/}} that allows users to search for 
archived copies of web pages (Mementos) across publicly available web archives, 
\item the Memento for Chrome\footnote{\url{https://bit.ly/memento-for-chrome}} and Memento for 
Firefox\footnote{\url{https://bit.ly/memento-for-firefox}} browser extensions that allow for browsing the past web,
\item the Redirect service\footnote{\url{http://timetravel.mementoweb.org/guide/api/\#memento-redirect}} 
that, for a given resource, redirects to the Memento created the closest to the desired past datetime, and
\item a variety of APIs\footnote{\url{http://timetravel.mementoweb.org/guide/api/}} aimed at providing Memento functionalities
to machines.
\end{itemize}
%
%
%
%
\begin{figure}[t!]
    \centering
    \begin{subfigure}[h]{0.5\textwidth}
        \includegraphics[scale=0.3]{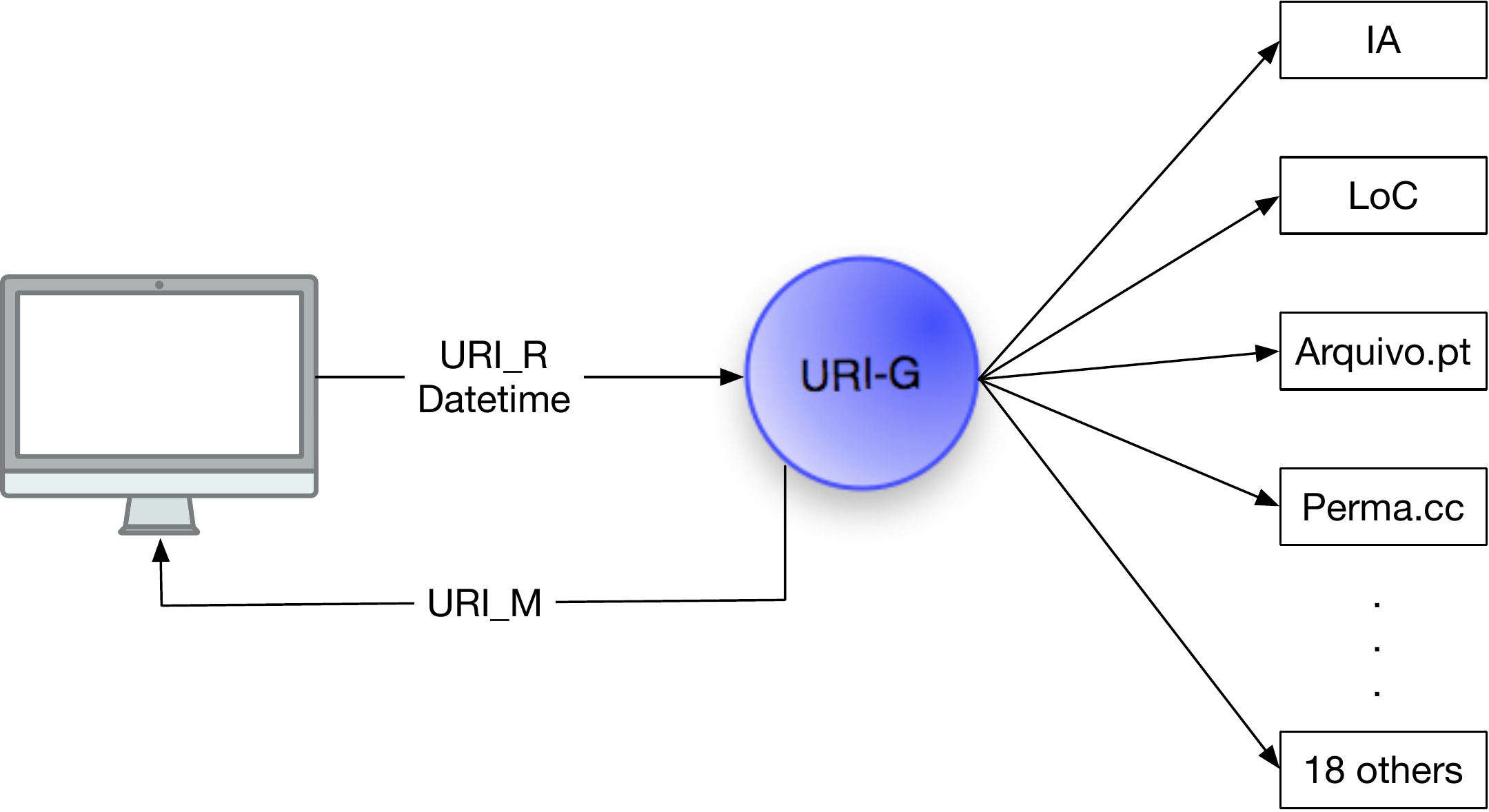}
        \caption{Previous implementation}
        \label{fig:ml_diagram1}
    \end{subfigure}
    \begin{subfigure}[h]{0.5\textwidth}
        \includegraphics[scale=0.3]{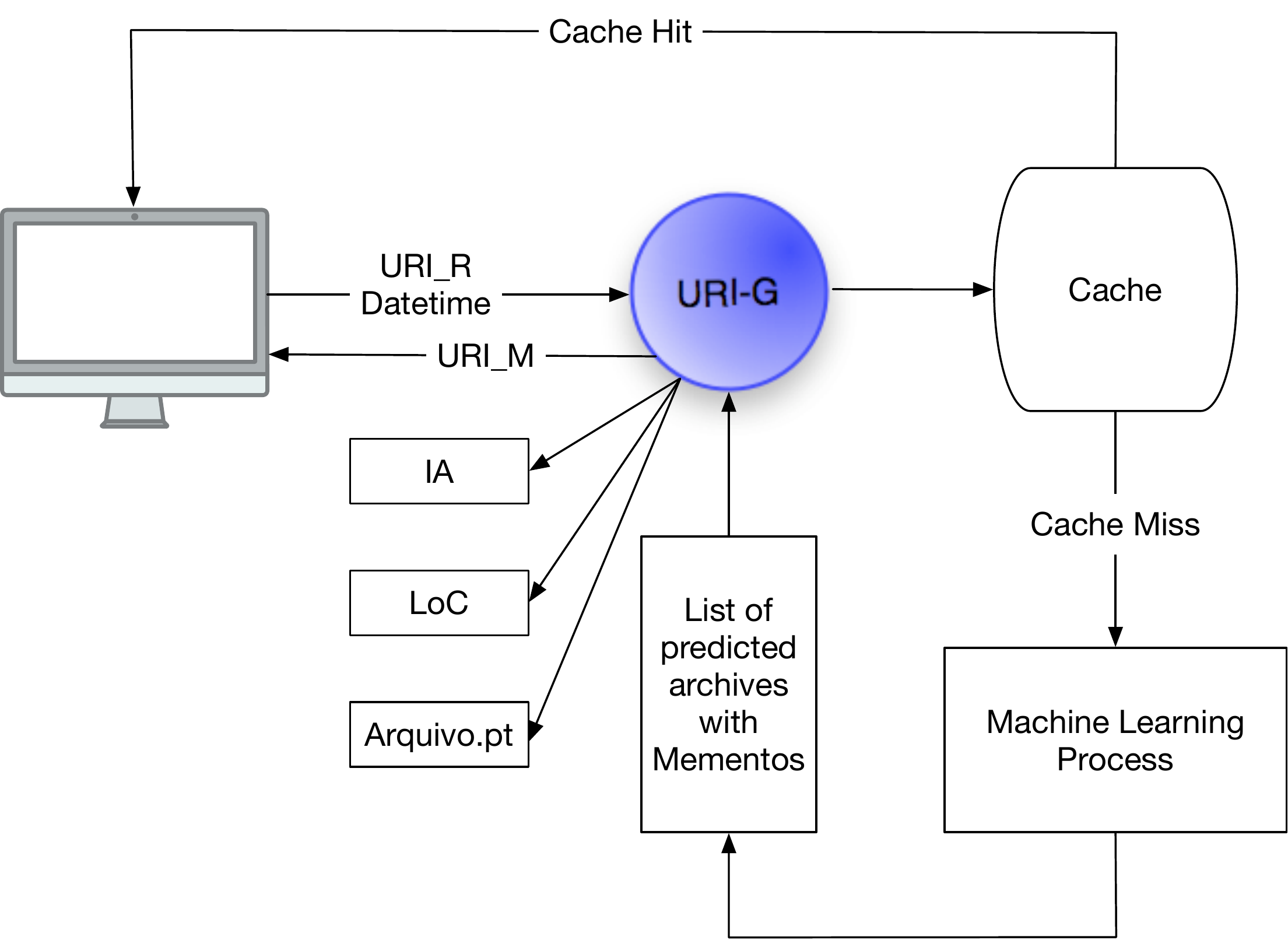}
        \caption{Current, optimized implementation}
        \label{fig:ml_diagram2}
    \end{subfigure}
    \caption{Memento Aggregator simplified structural view}
    \label{fig:ml_diagram}
\end{figure}
While these services allow users/machines to perform federated searches across more than $20$ publicly available web archives,
this functionality comes with challenges. In particular, with the number of available web archives growing, sending a request to 
each and every one of them for every single incoming Memento service request is not feasible. Particularly, overall 
response times and the load on the Memento as well as the individual web archive infrastructure are our major concerns.
Figure \ref{fig:ml_diagram1} shows a simplified view of the federated search behind the Memento services. The Memento 
Aggregator infrastructure (URI-G) receives a request with an original URI (URI-R) and a datetime in the past. The Aggregator
now contacts all web archives and asks for a Memento of the URI-R created as close as possible to the specified
datetime. The Aggregator receives URIs of Mementos (URI-Ms) from all archives that in fact have an archived copy available. 
This service therefore is only as fast as the slowest responding archive and, maybe more importantly, each of the archives 
responds to all requests, even if they do not have Mementos of the requested URI-R.
%
%

%
%
%
In order to optimize this federated search, we first established a cache system. Based on its resulting log data
we conducted a number of experiments to investigate the feasibility of simple machine learning classifiers to predict whether
a web archive has Mementos of a requested URI-R \cite{bornand:memento_routing}. The results were encouraging and hence we put the 
machine learning prediction process into production as well. Figure \ref{fig:ml_diagram2} shows a simplified view of the 
federated search process as it has been in production since $2016$. The Memento Aggregator infrastructure receives a request 
as shown in Figure \ref{fig:ml_diagram1} but now it first consults the cache. If a request for the same URI-R has been served
recently, its data is available in the cache and will therefore be served from cache (cache hit). This significantly
decreases the response time of the services. In case of a cache miss - no entry for the URI-R is found in the cache - the 
machine learning based process is started. As described in \cite{bornand:memento_routing}, it takes various simple characteristics
of the URI-R into account (length, number of tokens, top level domain, etc.) and predicts the archives in which Mementos are 
available. The Memento Aggregator infrastructure then contacts only these predicted archives and receives URI-Ms from those that
indeed have Mementos, which it eventually presents to the original requester. As a last step, the Aggregator contacts all other
web archives not predicted by the machine learning process and obtains URI-Ms from those that have Mementos and were
missed by our prediction. While this step still involves all archives and hence suffers from the same latency mentioned above, 
the requests are sent in batch mode and not optimized for speed, hence this process is much more friendly to the queried web 
archives. As a result of this step, the Aggregator eventually has the complete picture of available Mementos and populates the 
cache with this data. In practice, this means that responses orchestrated by the machine learning prediction are accurate but
may not be complete, where all responses from cache reflect accurate and complete data. 

In previous work \cite{bornand:memento_routing} we outlined the theoretical foundation of this process and presented initial 
experimental results that encouraged us to implement the concept presented here. In this work, we seek to evaluate both the 
cache and the machine learning based prediction. After almost two years in production, we can utilize the log files to offer 
insights into the following research questions:
\begin{enumerate}
\item How effective is the cache in terms of hit/miss rate? Does the ratio vary for different Memento services? Is the chosen 
cache freshness period appropriate?
\item How effective is the machine learning process in practice, compared to the baseline established in our previous work? 
How does the machine learning prediction perform in terms of keeping the number of false positives low? Do the models need to 
be retrained and if so how often?
\end{enumerate}
\section{Data collection}
To conduct our evaluation on the performance of the cache and the machine learning prediction, we studied the log
files of the above outlined four Memento services. For the machine learning evaluation we extracted data recorded between July 
$2017$ and October $2018$. This time frame is dictated by the last time we retrained the machine learning models (in July $2017$)
and by the time we conducted the experiment (starting in October $2018$). In total, the log files for the machine learning
evaluation contain $2,595,796$ entries. 
%
%

While the Memento services aggregate data from more than $20$ publicly available web archives, we limit our study to the $13$
archives listed in Table \ref{tab:web_archives}. These archives are all natively Memento-compliant and we have collected a
sufficient amount of usage data to train a machine learning prediction model for them. The reason the Internet Archive (IA),
the world's oldest and largest web archive, is not listed is because we do not maintain a prediction model for the IA. Its index
is so vast and so rapidly changing that we foresaw no model being able to accurately predict their holdings. Hence the IA is
excluded from our evaluations of recall and false positives.
\begin{table}[h!]
\caption{Evaluated web archives and their acronyms}
\begin{tabular}{|c|c|} \hline
\textbf{Acronym} & \textbf{Web Archive Name} \\ \hline \hline
archive.is & Archive.is \\ \hline
archive-it & Archive-It \\ \hline
ba & Bibliotheca Alexandrina Web Archive \\ \hline
blarchive & UK Web Archive \\ \hline
bsb & Bayerische Staatsbibliothek \\ \hline
gcwa & Canadian Archive \\ \hline
loc & Library of Congress \\ \hline
nli & National Library of Ireland \\ \hline
perma & Perma.cc \\ \hline
proni & The Public Record Office of Northern Ireland \\ \hline
pt & Arquivo.pt \\ \hline
swa & Stanford Web Archive \\ \hline
uknatarch & UK National Archives Web Archive \\ \hline
\end{tabular}
\label{tab:web_archives}
\end{table}
\section{Cache Evaluation}
There are many ways to implement \cite{podlipnig:web_cache_replacement} and optimize \cite{kelly:web_cache} cache systems but 
since our cache was for requests against web archive holdings rather than traditional web caches, we questioned whether these 
established methods were applicable. We therefore decided in favor of simplicity and implemented a binary cache for which
each request either results in a cache hit or a cache miss. We further decided to keep the cache records fresh for a period
of $30$ days. After a record has been in the cache for $30$ days, it is labeled ``stale'' and a request against the record 
results in a cache miss. A record is stale for another $30$ days before it ultimately is deleted.
\begin{figure}[t!]
\includegraphics[scale=0.3]{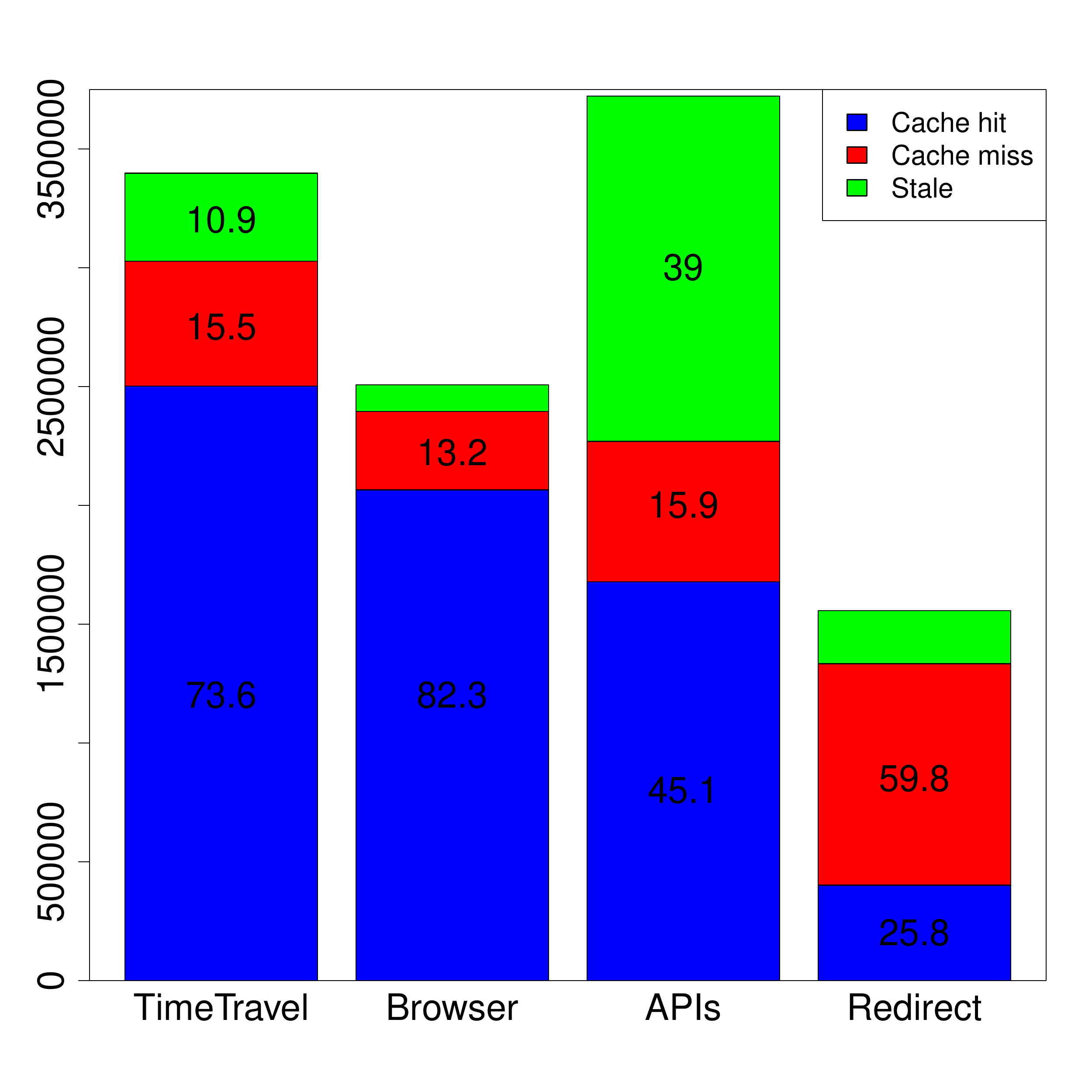}
\caption{Cache performance over four Memento services}
\label{fig:cache_rates}
\end{figure}
We were interested in investigating the performance of the cache over time in terms of cache hit vs. miss rate and the validity
of the $30$-day freshness period. Figure \ref{fig:cache_rates} shows the cache performance of four Memento services: 
TimeTravel, browser extension, APIs, and Redirect, each of which is represented by a bar from left to right. The blue portion of each 
bar indicates the cache hit ratio for the respective service, cache miss ratios are shown in red, and stale ratios in green. 
The overall height of a bar represents the total number of requests we have seen against the respective Memento service 
during the observed period of time.
We can make a few observations from Figure \ref{fig:cache_rates}. First, we see a high cache hit ratio for the 
TimeTravel service and the browser extension with $73.6\%$ and $82.3\%$ respectively. The API service with $45.1\%$ and especially
the Redirect service with $25.8\%$ show much lower numbers. This makes intuitive sense when considering the nature 
of the services. The former two are services aimed at human users and hence the requests are likely heavily influenced 
by popularity of the requested URIs. It is therefore reasonable to assume that URIs are being requested multiple times and
hence these services show a higher cache hit rate. The latter two services, on the other hand, are predominantly used by machines 
for batch requests, so likely include more random and not necessarily popular URIs, hence the lower hit rate. This seems especially
true for the Redirect service with a miss rate of $59.8\%$. It is interesting to observe a similar miss rate between the 
TimeTravel, browser extension, and API services but a significant stale rate for the APIs ($39\%$). We speculate that this is due to 
machines sending repeated requests of the same URIs, as it is frequently done in a variety of web archiving studies, and the 
experiments running longer than the cache entries are fresh.

We draw two main conclusions from Figure \ref{fig:cache_rates}. First, the cache works. On average, across all four
services, $59.4\%$ of requests are being served from cache, saving millions of requests to public web archives.
Second, while the $30$-day cache freshness period seems suitable for most services, we see reason to increase the period for the API
services in order to lower the ratio of stale requests.
%
%
%
\section{Machine Learning Evaluation}
\begin{figure}[t!]
\includegraphics[scale=0.3]{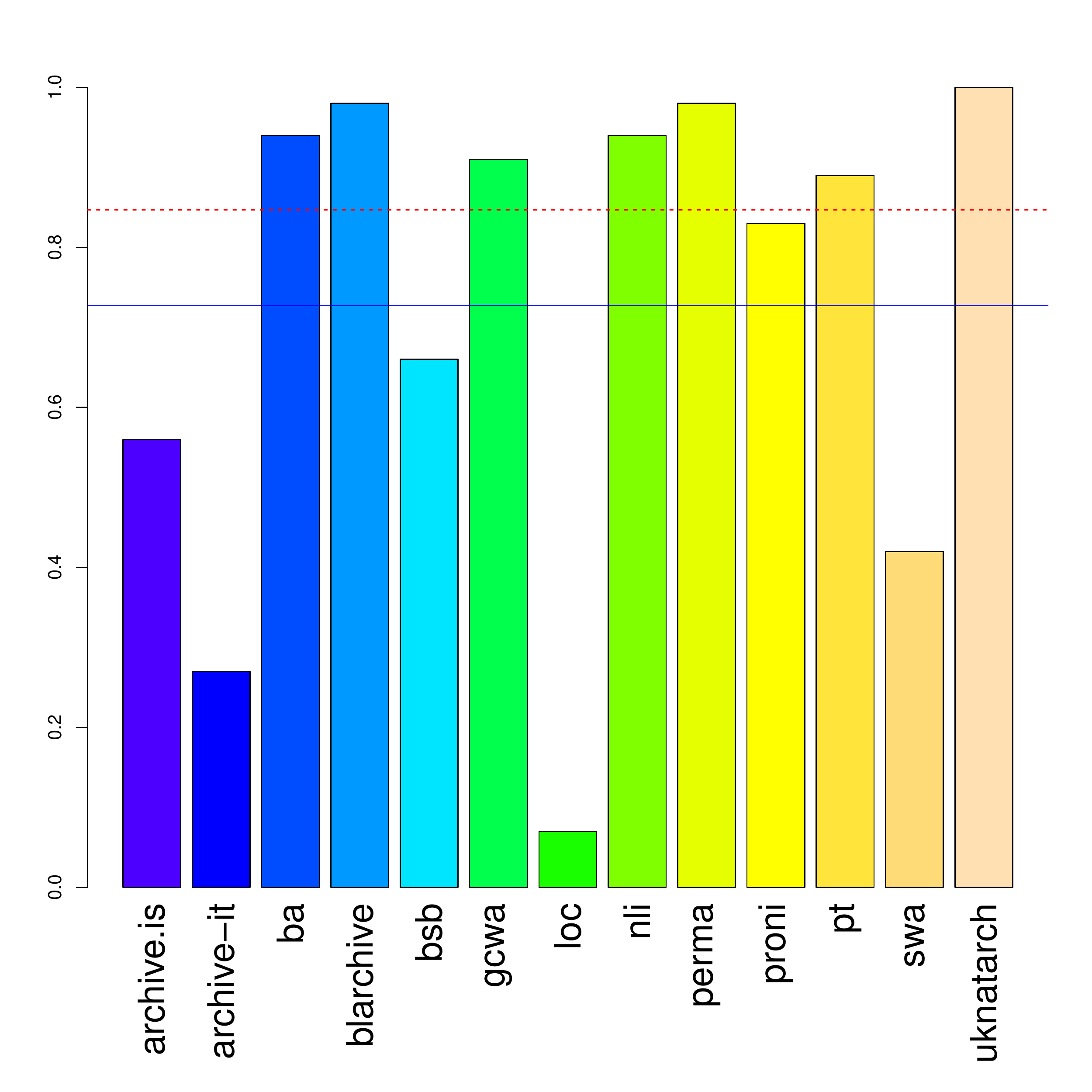}
\caption{Recall of web archives}
\label{fig:archives_recall}
\end{figure}

The experiment outlined in our previous work \cite{bornand:memento_routing} resulted in an observed recall value of $0.847$. We
take this value as the baseline and are motivated to investigate whether this baseline held true over time. 
Figure \ref{fig:archives_recall} visualizes all web archives and their computed average recall 
values\footnote{We compute recall values the same way as was done in \cite{bornand:memento_routing}: $recall=TP/(TP+FN)$,
where $TP$ represents true positives and $FN$ false negatives.}
over time, based on our log files. The figure further shows a dashed red horizontal line that intercepts with the y-axis at 
$0.847$ (the baseline) and a solid blue line at $0.727$, which is the average recall over all archives, based on our here 
computed data. 
We can observe a recall value for $8$ out of the $13$ web archives above or just below the baseline. The UK National Archives 
Web Archive has a perfect recall value of $1$ and the UK Web Archive as well as Perma.cc are almost perfect with a recall of 
$0.98$. 
These results are very encouraging and mean that our machine learning process is very accurate in predicting which archives 
hold a copy of a requested URI. 
However, Figure \ref{fig:archives_recall} also shows five archives with a recall value below the baseline and below our 
average. In particular, the recall values for Archive-It and the Library of Congress Web Archive are rather low with $0.27$ 
and $0.07$ respectively. These numbers confirm that our machine learning process is not operating on a satisfying level for
all web archives. One possible explanation for the low recall is that the web archive is very dynamic, meaning it frequently 
adds and removes entries from its index. Consequently, the data the machine learning process was trained on quickly does not
adequately represent the state of the archive anymore and hence the prediction mechanism results in misguided requests. Our
conclusion to this observation is the need for more frequent retraining processes for the archives with low recall numbers.

\begin{figure}[t!]
\includegraphics[scale=0.3]{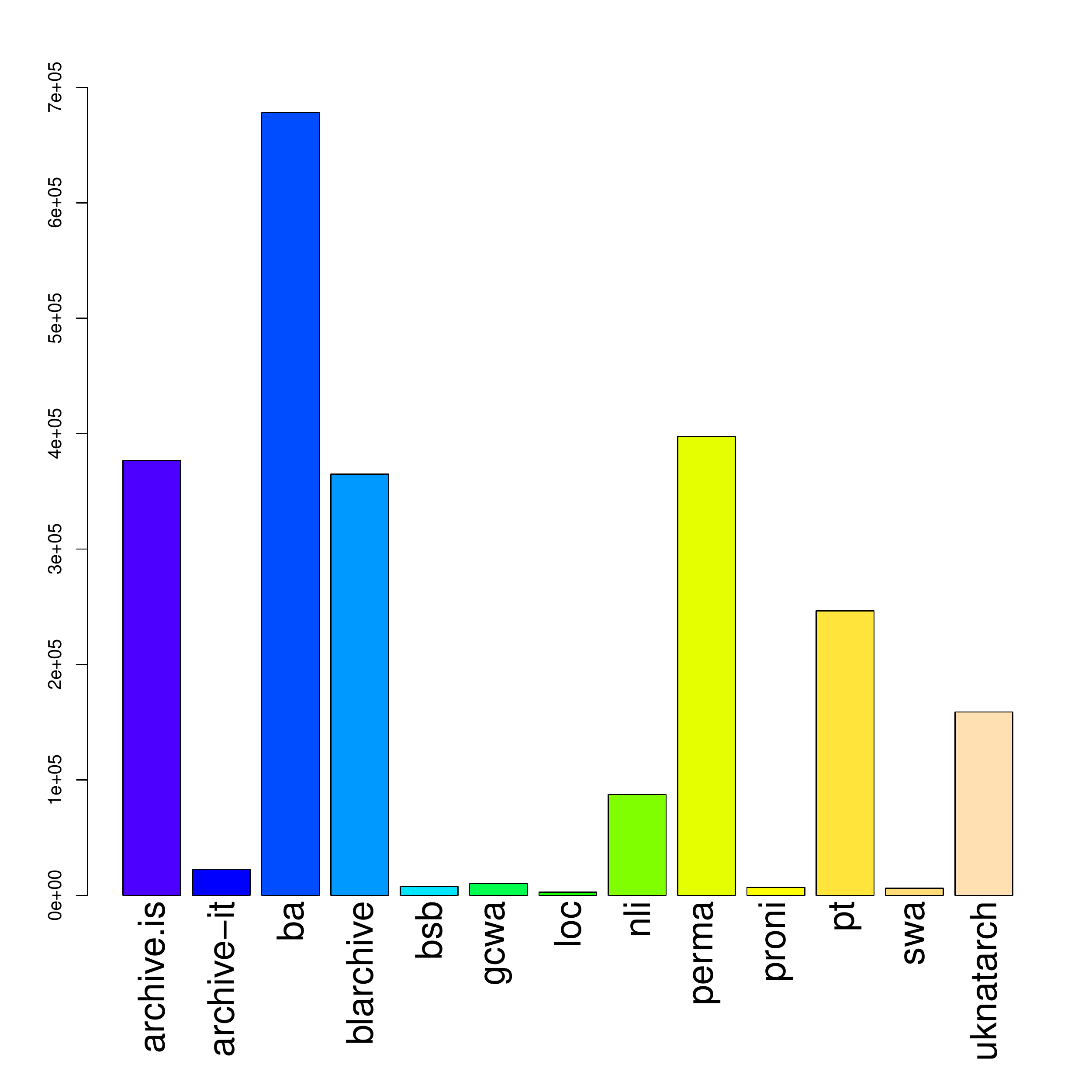}
\caption{False positives of web archives}
\label{fig:archives_fps}
\end{figure}

A different way to evaluate the performance of our machine learning prediction mechanism is to look at the number of false 
positives per archive. A false positive in this context occurs when the machine learning process predicts that a web archive 
has an archived copy of a requested URI where, in reality, it does not. False positives therefore result in requests that we 
seek to avoid as the contacted web archives return an empty result set.
Figure \ref{fig:archives_fps} displays the number of false positives for the $13$ archives. Immediately apparent is that the 
number for the Bibliotheca Alexandrina Web Archive is rather high ($677,999$). After consulting with staff from the archive, 
we were able to confirm that the archive was undergoing a lengthy infrastructure reorganization that frequently resulted in the 
web archive platform being unresponsive to requests from the Memento services, explaining the high number.
Other archives such as Archive.is ($376,835$), the UK Web Archive ($365,004$), and Perma.cc ($397,672$) also show high false
positive numbers. Possible reasons are the dynamic nature of a web archive such as Archive.is as mentioned earlier, technical
issues that for a period of time resulted in HTTP ``$404$ - File not found'' errors (UK Web Archive), and simply an imperfect trained
machine learning model, despite the high recall numbers (Perma.cc). The numbers shown in Figure \ref{fig:archives_fps} further
confirm our above mentioned conclusion for the need of a regular retraining of the machine learning models to accommodate for
changes in the archives that cause false positives.
\section{Dynamic Web Archives}
As mentioned, one possible explanation for low recall and false positives is the dynamic nature of web archives. Specifically, 
web archives quickly growing by continuously adding Mementos to their index or frequently deleting or preventing access 
to previously available Mementos. The former may be due to an institution's or even a country's archiving policies and 
the latter may be caused by take-down requests or other legal disputes. We investigated these dynamics, based on 
our log files, and how it affects our machine learning based prediction method. To this extent, we considered URI-Rs that have 
been requested multiple times from the Memento services and their requests (cache misses) caused the Aggregator every time to 
obtain the full picture of their Mementos distributed across various web archives.

\begin{figure}[th!]
\includegraphics[scale=0.3]{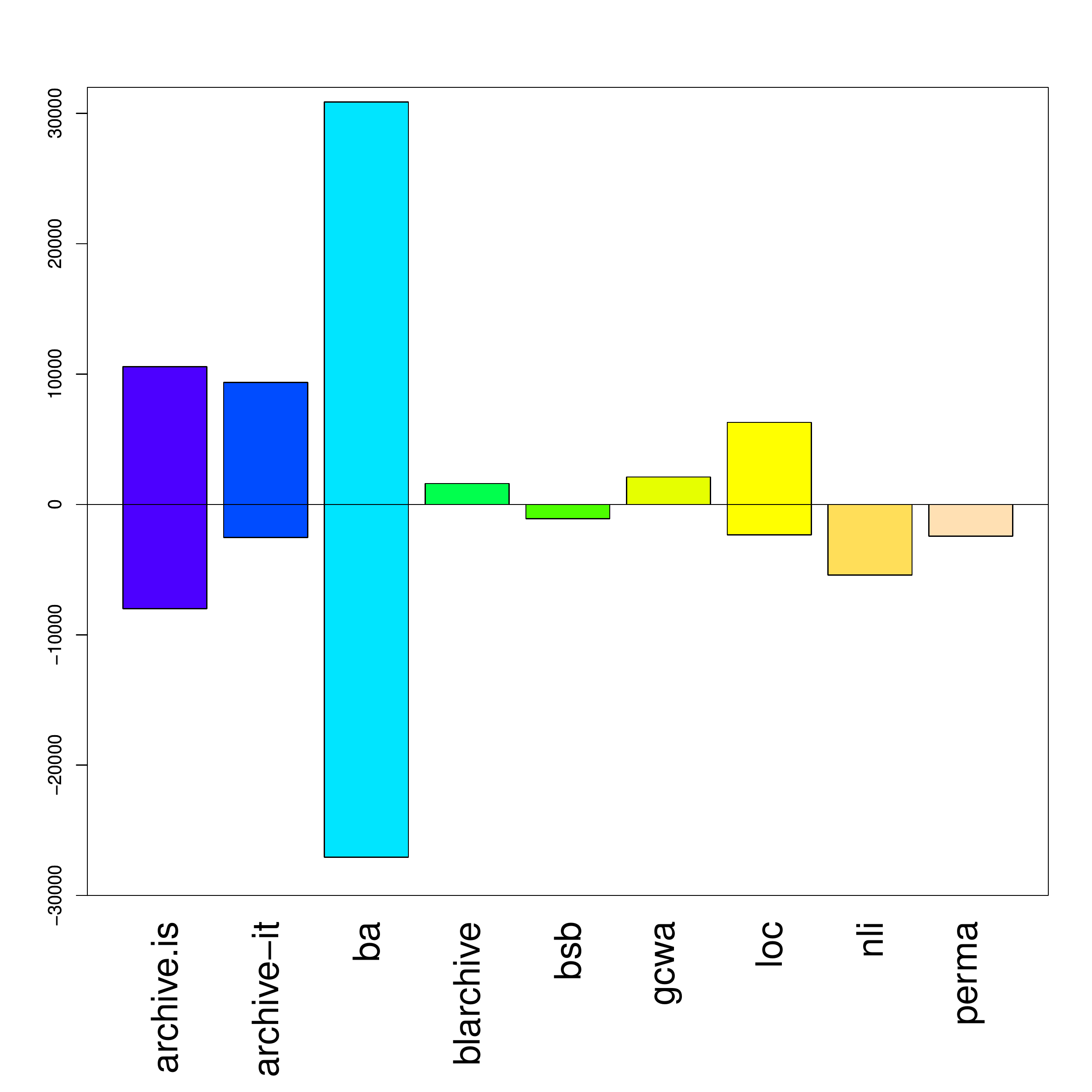}
\caption{Instances of archives being added/removed}
\label{fig:dynamics}
\end{figure}

Our log files revealed a total of $1,632,121$ such instances and we were able to determine that in $88.2\%$ of all cases the result
set of archives with available Mementos had not changed for multiple requests of the same URI-R. Changes in the result set that
do occur are split between $103,996$ cases ($6.4\%$) where archives were added and $88,727$ cases ($5.4\%$) where archives were 
removed. 
The IA leads the statistics; it was added to a result set more than $44$ thousand times and removed from one more than $27$ thousand
times. Figure \ref{fig:dynamics} displays the amount of times archives that are subject to our machine learning prediction
were added/removed from a result set. The figure confirms the dynamic character of some of the archives. In particular, Archive.is, 
Archive-It, and the Library of Congress were added to and removed from numerous result sets. The large number of instances where
the Bibliotheca Alexandrina Web Archive was added and removed can be attributed to a large number of transient errors the Memento
services observed, likely due to the infrastructure restructuring mentioned above.
The results presented in Figure \ref{fig:dynamics} further strengthen our argument that the machine learning retraining is essential 
to keep up with the evolution of the archiving landscape and continuously serve results with high 
recall and as few as possible false positives.
\section{Conclusion}
In this study we report on an evaluation of the effectiveness of our cache system and machine learning prediction - two
optimization mechanisms for Memento Aggregator-based services. We find the cache to be very effective, especially for human-driven 
services, where we observe a cache hit rate of more than $80\%$. However, there clearly is room for improvement for machine-driven 
services. We further conclude that the overall, machine learning prediction models operate at an acceptable recall level of $0.727$ 
but we also show clear indicators that support the need to retrain our models on a regular basis in order to keep up with the 
changing web archive landscape and to further lower the false positive rates per archive.
Going forward, we are interested in exploring the performance of machine learning models that are based on archival holdings and 
not on our recorded usage data. In addition, we are experimenting with neural network classifiers for the prediction task and first 
results are very promising yet too premature to be reported here.
%
%
%
%
%
%

%
\end{document}